\documentclass[a4paper,12pt]{article}
\title{\bf 
Glueball Masses from Supergravity with Flavor 
}
\author{ Kenji Suzuki \footnote{ksuzuki@sofia.phys.ocha.ac.jp} }

\date{}

\begin{document}
\maketitle

%\author{ Kenji Suzuki \\
\begin{center}
\vspace{5mm}
{\it 
  Department of Physics, Ochanomizu University, \\
 Otsuka, Bunkyou-ku, Tokyo 112-8610, Japan\\

\vspace{3mm}

  Institute of Quantum Science, \\
College of Science and Technology, 
Nihon University, \\
 Chiyoda-ku, Tokyo 101-8308, Japan
%\email{ksuzuki@sofia.phys.ocha.ac.jp}
}
\end{center}

\begin{abstract}
We study the glueballs
in a four-dimensional ${\cal N}=2$ super Yang-Mills theory with fundamental matters 
in terms of the supergravity dual.
The supergravity background is constructed by $N$ D3 brane with a probe D7 brane.
We numerically compute the glueball masses for $0^{++}$ and $1^{--}$
in the background. 
We find that the mass ratio of $0^{++}$ glueballs is mostly 
in agreement with the lattice calculations.  
We compare the glueball masses with the meson masses. 
The mass ratio $M_{GB}/M_{meson}$ is calculated with $1.4$.
If the mass $M_{GB}$ is set to $1.6 GeV$, 
the meson mass $M_{meson}$ is given with $1.1 GeV$
\end{abstract}

%\preprint{hep-th/0411076}

%\keywords{AdS-CFT Correspondence, D-branes, Supersymmetry and Duality }

\newpage

%\begin{document}
\section{Introduction}
The AdS/CFT correspondence~\cite{mal,wit,gub,ah} is a duality between 
the supersymmetric gauge theories in the large $N$ strong-coupling limit 
and the string theories on the $AdS$ backgrounds.
As the generalization of the duality,
the supergravity duals to the gauge theories with less supersymmetries 
have been investigated~\cite{it,wit2,gi,ks,mn,ba}.
According to the method~\cite{mal2},
the behaviours of Wilson loops for the gauge theories with less supersymmetries
%for the strong coupling limit
%can be  
have been calculated
in the dual gravity  backgrounds~\cite{br,lo,mat,ap},
which indicate that the gauge theories are in confinement phase.
The mass spectra for the $0^{++}$ glueball in the dual background
can be obtained by solving the wave equations of motion
for the dilaton~\cite{wit2,cs,mj,mi,bro,bb,ca,am}, 
which are in agreement with the lattice calculations~\cite{te,mo}.

In order to obtain the supergravity dual of a four-dimensional ${\cal N}=2$ 
$SU(N)$ super Yang-Mills theory with $N_f$ hypermultiplets 
%in the fundamental representation, 
Karch and  Katz consider the background 
embedded $N_f$ D7-branes into $N$ D3-branes~\cite{kk,kkw}.
The masses of the hypermultipletes (quarks) is given with the distance
between the D3-branes and the D7-brane.
The mesons, which correspond to open string excitations of the D7-brane,
can be calculated in the background~\cite{kmmw}.
The ${\cal{N}}=1$ cases are also studied~\cite{npr,blzv}.
In the non-supersymmetric cases,
the chiral symmetry breaking by the quark condensate in terms of the gravity dual 
is studied in the three-dimensional QCD~\cite{ba2} and 
in the four-dimensional QCD~\cite{kmmw2}.
However, the glueballs in the dual background with hypermultiplets 
is not investigated .

%・meson spec
%・non-susy D3/D7
%・non-susy D4/D6
%・chiral sym bra
%・beond probe appli
%・GMOR,  Regge

%intro:
%・AdS/CFT corr、Gauge/Gravity
%・Wilson loop
%・Glueball
%・N=1 KS, MN
%・Flavor, meson

The purpose of this paper is to study the glueballs 
in a four-dimensional ${\cal N}=2$ super Yang-Mills theory with fundamental matters 
in terms of the supergravity dual. 
We calculate the glueball masses and compare them with the meson masses. 
We show that the glueball masses is proportional to the mass of the 
fundamental matter.
The mass spectrum in the background are in agreement with the lattice calculations.

The organization of the paper is as follows.
In section 2, we will review the calculations of the meson spectrum 
in the supergravity background constructed by $N$ D3 brane with a probe D7 brane
discussed in~\cite{kmmw}.
In section 3, we will solve the wave equation of the scalar fields in the background
numerically to obtain the $0^{++}$ glueball mass spectrum.
We compare the results with the lattice computations and 
evaluate the mass ratio of the glueball and the meson.
In section 4, we will solve the equation of the RR 2-form numerically 
to obtain the $1^{--}$ glueball mass spectrum.
We compare the spectrum with the calculation~\cite{ca}. 

%\newpage

\section{Mass spectrum for meson}
In this section,
we review the derivation of the mass spectrum for meson
using the supergravity dual of a four dimensional supersymmetric Yang Mills theory
with fundamental matter.

We first consider a D7-brane probe in the supergravity background dual to $N$ D3-branes
represented by the array 
\begin{center}
%\begin{eqnarray}
D3: \ t \ 1 \ 2 \ 3 \, - \, - \, - \, - \ - \ -  \\
 \ \ \ D7: \ t \ 1 \ 2 \ 3 \ 4 \ 5 \ 6 \ 7 \ - \ - \ . 
%\end{eqnarray}
\end{center}
%\begin{equation}
%ds^2 = f^{-1/2} ( -d t^2 + d \vec{x}^2) 
%    + f^{1/2} (d r^2 + r^2 d \Omega^2_5 ),
%\end{equation}
The metric is given by
\begin{eqnarray}
 ds^2 = f^{-1/2} (- dt^2 + d \vec{x}^2)
 + f^{1/2} (d\lambda^2 + \lambda^2 d \Omega_3^2+dx_8^2+dx_9^2),
\end{eqnarray}
where $\vec{x} = (x_1,x_2,x_3)$,  and
\begin{eqnarray}
&& f = 1 + \frac{\alpha'^2 R^4}{r^4} , \\
&& R^4 = 4\pi g N ,\\
&& e^{\phi} =g ,\\
&& r^2 = \lambda^2+x_8^2+x_9^2.
\end{eqnarray} 
In the decoupling limit $\alpha'\to 0$, the metric becomes
\begin{eqnarray}
ds^2 / \alpha' = \bigg(\frac{u}{R}\bigg)^2 (-d t^2 + d \vec{x}^2 ) 
    + \bigg(\frac{R}{u}\bigg)^2 (d\rho^2 + \rho^2 d \Omega_3^2
    + dX_8^2 + dX_9^2),
\end{eqnarray}
where
\begin{eqnarray}
&& u = \frac{r}{\alpha'}, \ \ \rho = \frac{\lambda}{\alpha'}, \ \
X_8 = \frac{x_8}{\alpha'}, \ \ X_9 = \frac{x_9}{\alpha'}\ \ \ : fixed \\
&&u^2 = \rho^2 + X_8^2+X_9^2.
\end{eqnarray}

We next consider the meson spectrum in the background. 
The meson correspond to open string excitations of the D7-brane.
The DBI action for the D7 brane in this background, 
\begin{eqnarray}
 S_{D7}&=&-T_7\int d\sigma^8e^{-\phi}\sqrt{-det g} \nonumber \\
% &=&-T_7\int d\sigma^8\epsilon_3\rho^3e^{-\phi} 
%\sqrt{1+\frac{g^{ab}R^2}{u^2}(\partial_aX_8\partial_bX_8+\partial_aX_9\partial_bX_9)}\\
 &\sim& -T_7\int d\sigma^8\epsilon_3\rho^3e^{-\phi} 
\bigg\{1+\frac{g^{ab}R^2}{2u^2}(\partial_aX_8\partial_bX_8+\partial_aX_9\partial_bX_9)
\bigg\},
% &\sim& -T_7\int d\sigma^8\epsilon_3\rho^3e^{-\phi} 
%(1+\alpha'^2\frac{g^{ab}R^2}{2u^2}\partial_a\delta\partial_b\delta)
\end{eqnarray}
where $g_{ab}$ is the induced metric on the D7 brane, $\epsilon_3$ is the volume of
the three sphere, and $T_7=1/(2\pi)^7g\alpha'^4$ is the D7 tension.
To obtain the mass spectrum of the meson, we take the form as 
\begin{eqnarray}
 &&X_9 + iX_8 = m + \delta(\rho)e^{ik_i x^i}, \\ 
 &&m_q = m/2\pi, \ \ M^2 = -k^2, \ \ (i = 0,1,2,3) ,
\end{eqnarray}
where $\delta(\rho)$ is a small fluctuation. 
The mass of the quark $m_q$ is proportional to the distance $m$
between the D3-branes and the D7-brane.
$M$ is the meson mass.
The equation of motion becomes
\begin{eqnarray}
 \partial^2_\rho \delta(\rho)+\frac{3}{\rho}\partial_\rho \delta(\rho)
 + \frac{M^2R^4}{(\rho^2+m^2)^2}\delta(\rho)=0.
\end{eqnarray}
The solution can be written as
\begin{eqnarray}
 &&\delta(\rho) = \frac{A}{\rho^2+m^2}F(-n-1,\ -n,\ 2;\ -\rho^2/m^2)
\end{eqnarray}
where $F$ is the hypergeometric function.
We finally obtain the mass spectrum for meson to be 
\begin{eqnarray}
 &&M = \frac{2m}{R^2}\sqrt{(n+1)(n+2)},\ \ \ 
(n=0,1,2,\cdots).\label{meson}
\end{eqnarray}
Therefore the mass of meson in the ${\cal N}=2$ cases
is proportional to the quark mass in the UV region.
It is known that in the non-supersymmetric cases 
the mass of meson is proportional to the square root of the quark mass,
which relation is in agreement with the Gell-Mass-Oakes-Renner 
(GMOR) relation~\cite{kmmw2} 

%\newpage

\section{Mass spectrum for $0^{++}$ glueballs}
In this section,
we consider the mass spectrum for $0^{++}$ glueball 
using the supergravity dual description. 

We take the location of the D7-brane at $X_8^2 + X_9^2 = m^2$.
The induced metric is 
\begin{eqnarray}
ds^2 / \alpha' = \frac{\rho^2+m^2}{R^2} (-d t^2 + d \vec{x}^2 ) 
    + \frac{R^2}{\rho^2+m^2} (d\rho^2 + \rho^2 d \Omega_3^2 ),
\end{eqnarray}
discussed in the previous section.
In order to compare the glueball masses with the meson masses at the same time, 
it is necessary to consider the duals of glueballs (closed strings) 
in the same background as in the case of the mesons. 
We suggest that the mass spectrum for $0^{++}$ glueball
can be obtained by solving the wave equation of the scalar fields 
in the supergravity dual background.
The wave equation is
\begin{eqnarray}
 \partial_{\mu}e^{-2\phi}\sqrt{-g}g^{\mu\nu}\partial_\nu \Phi = 0.
\end{eqnarray}
We take the form $\Phi = \phi(\rho) e^{ik_i x^i} \ \ (i = 0,1,2,3)$.
The equation becomes
\begin{eqnarray}
 \partial_\rho \{\rho^3(\rho^2 + m^2) \partial_{\rho} \phi\}
+\frac{\rho^3}{\rho^2+m^2} M^2R^4 \phi = 0 \\ 
 M^2 = -k^2, \ 
\end{eqnarray}
where $M$ is the glueball mass.
%If $m \neq 0$,% 
Using the dimensionless variable $\rho' \equiv \rho/m$
with non-zero $m$,
the equation becomes
\begin{eqnarray}
 \partial_{\rho'} \{{\rho'}^3({\rho'}^2 + 1) \partial_{\rho'} \phi\}
+\frac{{\rho'}^3}{{\rho'}^2+1} \frac{M^2R^4}{m^2} \phi = 0.
\end{eqnarray}
The equation has asymptotic solution of the form $\phi \sim c/{\rho'}^4$,
where $c$ is arbitrary constants.
We solve the equation by shooting method~\cite{cs}, 
integrating from large $\rho' >>1 \ (i.e. \ \rho >> m)$
to $\rho' = 0$. 
We show the results in table 1.
We note that the masses are proportional to the quark mass.
The lightest mass is
\begin{eqnarray}
 M \sim 4.1 \frac{m}{R^2}.\label{gb}
\end{eqnarray}
In table 2, we compare our results with lattice calculations and 
with the results in the other backgrounds.
We find that our results are mostly agreement with the results of lattice calculations
and others.

%\TABLE{
\begin{table}
\begin{center}
\begin{tabular}{|c|c|} \hline
  $MR^2/m$        &D3/D7  \\ \hline
  $0^{++}$   & 4.1  \\ \hline
  $0^{++*}$  & 6.2  \\ \hline
  $0^{++**}$ & 8.3  \\ \hline
  $0^{++***}$ & 10.4  \\ \hline
  $0^{++****}$ & 12.6  \\ \hline
\end{tabular}
\caption{$0^{++}$ glueball masses}
%}
\end{center}
\end{table}

We next consider to compare the glueball masses with the meson masses.
The mass ratio of the lowest $0^{++}$ glueball state (\ref{gb})
and the lowest meson state (\ref{meson}) is written as
\begin{eqnarray}
\frac{M_{0++}}{M_{meson}} \sim \frac{4.1}{2\sqrt{2}} \sim1.4 ,
\end{eqnarray}
which is independent to the quark mass $m_q= \frac{m}{2\pi}$.
We note that the mass ratio of the glueballs and of the mesons are given by
the the square root of ratio of the tension $T_{adj}$ and the tension $T_{QCD}$~\cite{zsv},
namely 
\begin{eqnarray}
\frac{M_{0++}}{M_{meson}} = \sqrt{\frac{T_{adj}}{T_{QCD}}}
 = \sqrt{\frac{2N^2}{N^2-1}} \sim 1.4 
\end{eqnarray}
in large $N$ limit, which is consistent with our results.
If the mass $M_{GB}$ is set to $1.6 GeV$, 
the meson mass $M_{meson}$ is given with $1.1 GeV$.

%\newpage

%\TABLE{
\begin{table}
\begin{center}
\begin{tabular}{|c|c|c|c|c|c|} \hline
  $M$  & D3/D7  & YM${}^*$~\cite{ap} & KS~\cite{ca} & ${\cal N}=0$~\cite{cs} 
& Lattice~\cite{te,mo} \\ \hline
  $0^{++}$   &1.6(input) &1.6(input) &1.6(input) &1.6(input) &1.6$\pm$0.15 \\ \hline
  $0^{++*}$  &2.4        &2.4 &2.9        &2.6 &2.48$\pm0.23$ \\ \hline
  $0^{++**}$ &3.2        &3.1 &?          &3.5 &? \\ \hline
\end{tabular}
\caption{The comparison of $0^{++}$ glueball masses in GeV}
%}
\end{center}
\end{table}

%\newpage

\section{Mass spectrum for $1^{--}$ glueballs}
In this section, we consider the $1^{--}$ glueballs in the supergravity description. 
The operators of $1^{--}$ glueball are dual to the antisymmetric tensor fields.
The equation for the anti-symmetric tensor field $A_{\mu\nu}$ is given by
\begin{eqnarray}
 \partial_{\mu}(\sqrt{-g}g^{\mu'\mu}g^{\mu'_1\mu_1}g^{\mu'_2\mu_2}
\partial_{[\mu'} A_{\mu'_1\mu'_2]}) = 0.
\end{eqnarray}
We take the form $A_{\mu\nu} = h(\rho) e^{ik_ix^i} \ \ (i = 0,1,2,3)$. 
The equation becomes 
\begin{eqnarray}
 \partial_{\rho}(\frac{\rho^3}{\rho^2+m^2} \partial_{\rho} h)
+\frac{1}{3}\frac{\rho^3}{(\rho^2+m^2)^3} M^2R^4 h = 0,
\end{eqnarray}
where $M^2 = -k^2$. 
The equation has asymptotic solutions of the form $h \sim c \ln \rho$,
where $c$ is arbitrary constant.
We solve the equation by shooting method, integrating from large $\rho'= \rho/m >> 1$
to $\rho' = 0$.
We show the results in table 3 and table 4.
We note that the masses are proportional to the quark mass.
The lowest mass is written as 
\begin{eqnarray*}
 M \sim 4.7 \frac{m}{R^2}.
\end{eqnarray*}
\begin{table}
\begin{center}
%\TABLE{
\begin{tabular}{|c|c|} \hline
  $MR^2/m$        &D3/D7  \\ \hline
  $1^{--}$   & 4.7  \\ \hline
  $1^{--*}$  & 8.9  \\ \hline
  $1^{--**}$ & 12.6  \\ \hline
  $1^{--***}$ & 16.3  \\ \hline
  $1^{--****}$ & 19.8  \\ \hline
\end{tabular}
\caption{$1^{--}$ glueball masses}
%}
\end{center}
\end{table}

We consider the mass ratio of the lowest $0^{++}$ glueball and 
the lowest $1^{--}$ glueball.
We compare it in our results with 
in the results in the ${\cal N}=1$ Klebanov-Strassler background~\cite{ca}.
\begin{eqnarray*}
&&\bigg(\frac{M_{1--}}{M_{0++}}\bigg)_{D3/D7} = 1.15 \\
&&\bigg(\frac{M_{1--}}{M_{0++}}\bigg)_{KS} = 1.20 
\end{eqnarray*}
We find that our results are mostly in agreement with the results 
in the Klebanov-Strassler background.
\begin{table}
\begin{center}
%\TABLE{
\begin{tabular}{|c|c|c|} \hline
  $M$        & D3/D7     &KS~\cite{ca}  \\ \hline
  $1^{--}$   &1.8 &1.9 \\ \hline
  $1^{--*}$  &3.4 &3.3             \\ \hline
  $1^{--**}$ &5.2 &?             \\ \hline
\end{tabular}
\caption{The comparison of $1^{--}$ glueball masses in GeV \ (input $0^{++}=1.6 GeV$).}
%}
\end{center}
\end{table}

\section{Conclusions}

We study the glueball 
in the four-dimensional ${\cal N}=2$ super Yang-Mills with fundamental matters.
We numerically compute the glueball masses for $0^{++}$ and $1^{--}$.
We find that the masses are proportional to the quark mass.
The mass ratio of $0^{++}$ glueballs is mostly in agreement 
with the results of lattice calculations.  
The mass ratio of $1^{--}$ glueballs is mostly in agreement 
with the results in the ${\cal N}=1$ Klebanov-Strassler background.

We evaluate the mass ratio of the glueball and the meson, 
which is written as $M_{GB}/M_{meson}$ $\sim 1.4$.
The mass ratio of the glueballs and of the mesons are given by
the the square root of ratio of the tension $T_{adj}$ and the tension $T_{QCD}$,
namely ,
\begin{eqnarray}
\frac{M_{0++}}{M_{meson}} = \sqrt{\frac{T_{adj}}{T_{QCD}}}
 = \sqrt{\frac{2N^2}{N^2-1}} \sim 1.4 
\end{eqnarray}
in large $N$ limit, which is consistent with our results.

It would be interesting in calculating the $2^{++}$ glueball masses 
in the dual background in order to study the Regge trajectory of the glueball masses.
It was already shown that the Regge trajectory is linear 
in the Klebanov-Strassler background~\cite{ac}.

We are also interested in studying the glueballs, 
especially, in the non-supersymmetric cases with massless quarks,  
which indicate that the chiral symmetry is broken by the quark condensate.

\end{document}